\documentclass[12pt]{article}
\usepackage{a4wide}
\usepackage{amssymb}
\begin{document}
\newcommand{\be} {\begin{equation}}
\newcommand{\ee} {\end{equation}}
\newcommand{\ba} {\begin{eqnarray}}
\newcommand{\ea} {\end{eqnarray}}
\newcommand{\e} \epsilon
\newcommand{\la} \lambda
\newcommand{\La} \Lambda 
%%%%%%%%%%%%%%%%%%%%%grant2%%%%%%%%%%%%%%%%
\author{B. Boisseau\thanks{E-mail :
boisseau@phys.univ-tours.fr}\\
\small Laboratoire de Math\'ematiques et Physique Th\'eorique\\
\small CNRS/UMR 6083, Universit\'e Fran\c{c}ois Rabelais\\
\small Facult\'e des Sciences et Techniques\\
\small Parc de Grandmont 37200 TOURS, France}
%%%%%%%%%%%%%%%%%%%%%%%%%%%%%%%%%%%%%%%%%%%%%%%%%
\title{\bf Vortex in a weakly relativistic Bose gas at zero temperature and 
relativistic  fluid approximation }
\date{}
\maketitle
%%%%%%%%%%%%%%%%%%%%%GRANT2%%%%%%%%%%%%%%%%%%%%%%%%
\begin{abstract}

The Bogoliubov procedure in quantum field theory is used to describe 
a relativistic almost ideal Bose gas at zero temperature. 
Special attention is given to the study of a vortex. The radius of the vortex 
in the field description is compared to that obtained in the relativistic fluid approximation. 
The Kelvin waves are studied and, for long wavelengths, the dispersion relation is 
obtained by an asymptotic matching method and compared with the non relativistic result.


\end{abstract}

%%%%%%%%%%%%%%%%%%%%%%%%%%%%%%%%%%%%%%%%%%%%%%%%%%%%%%%

\section{Introduction}

%%%%%%%%%%%%%%%%%%%%%%%%%%%%%%%%%%%%%%%%%%%%%%%%%%%%%%%

An interesting issue in the study of relativistic vortices involves 
the similarities and differences between global strings and relativistic 
superfluid vortices \cite{BY5,BY8,gradw,BY6,BY10}.

This paper deals with a relativistic treatment of the almost ideal Bose gas at zero temperature
and its approximation as a relativistic fluid. More specifically we study a cylindrical vortex 
and its Kelvin waves.  

We know that an almost ideal Bose gas at zero temperature is well described 
by the Gross-Pitaevskii equation \cite{gros61,pit61,gros63} which governs the 
evolution of the macroscopic wave function caracterising the condensate. This suggests (Sec.2) to start with a quantum
field whose potential $m^2 \phi^{*}\phi+\lambda(\phi^{*}\phi)^2$
does not exhibit spontaneous symmetry breaking, in order to represent 
weakly interacting relativistic Bose particles. By the 
Bogoliubov procedure we show that the classical field equation can be 
interpreted as a ``relativistic Gross-Pitaevskii equation''. In order to apply 
the Bogoliubov procedure we have to verify that the mode, in which the bosons 
condense, realizes the minimum of energy with the constraint that the 
conserved charge is fixed by the number of particles. This is done in Sec.3.

This preliminary study makes things very similar to the non relativistic 
counterpart (cf. for example \cite{dalf}) and allows us to interpret 
the classical field equation as 
representing the motion of the condensate of an almost ideal gas at zero 
temperature. 

In Sec.4 we establish  the approximation of the field equations by a perfect
isentropic irrotational fluid in the formulation of Lichnerowicz \cite{lich}. 
In Sec.5 the cylindrical stationary solution(that is a vortex) is studied. 
We begin with the exact solution in the approximate fluid theory and
compare with an asymptotic solution of the exact theory. 

In Sec.6 we study the Kelvin waves. The oscillations of a rectilinear vortex were initially studied 
by Lord Kelvin \cite{thomson} in the context of the classical fluid. 
In Bose condensate theory the oscillations of a non relativistic quantum vortex 
were first discussed by Pitaevskii \cite{pit61} who found in the long wavelength limit the dispersion relation:
\be
\label{101}
\omega=\frac{q^2}{2m}\ln{\frac{1}{qr_0}}
\ee
where $r_0$ is the core radius.
This result is refined by Grant \cite{grant} who has found:
\be
\label{102}
\omega=\frac{q^2}{2m}\{\ln{\frac{2}{qr_0}}-\gamma-0.115\}
\ee
where $\gamma$ is the Euler constant. Davis and Shellard \cite{BY5} have shown that a spinning global string 
behaves like a vortex in a superfluid medium. The dispersion relation of the Kelvin modes of such a vortex 
was given by Ben-Ya'acov \cite{ben4}. 
In the non relativistic limit, he obtains
\be
\label{103}
\omega=\frac{q^2}{2m}\{\ln{\frac{1}{q\epsilon}}-\gamma+1.5\}
\ee
where $\epsilon$ is a cutoff parameter which depends on the core radius parameter $\delta$ of the string by
\be
\label{103'}
\ln{\frac{\epsilon}{\delta}}=1.615-\ln 2.
\ee
This last formula is obtained for a vortex ring. Comparison of (\ref{102}) and (\ref{103}) yields exactly the same 
formula if $r_0=\delta$.

In this paper we cannot follow the method of \cite{ben4} which uses 
explicitly an equation of motion of the vortices based 
on the spontaneous breaking symmetry of the potential. 
We shall rather transpose the method of the matched-asymptotic-expansions of 
Roberts and Grant \cite{robertsgrant, grant} to the relativistic field equations. 
The result is
\be
\label{104}
\omega=\sqrt{1-v_0^2}\frac{q^2}{2m}\{\ln{\frac{2}{qr_0}}-\gamma-0.115\}
\ee 
where $v_0$ is the velocity at $r=r_0$ of the classical relativistic fluid associated with the condensed bosons.
 Finally in Sec.7 we discuss some of our results, in particular we compare (\ref{104}) with the dispersion 
relation of the Kelvin modes of a global string.

In this article the signature of the metric is $(-+++)$ and $c=\hbar=1$.

%%%%%%%%%%%%%%%%%%%%%%%%%%%%%%%%%%%%%%%%%%%%%%%%%%%%%%%%%%%%%%%%%%

\section{Classical field equation and Bogoliubov procedure}

%%%%%%%%%%%%%%%%%%%%%%%%%%%%%%%%%%%%%%%%%%%%%%%%%%%%%%%%%%%%%%%%%%

Let us consider a phenomenological system of electrically neutral bosons 
described by a complex scalar field. It is worth noting that a complex field (non Hermitian operator) can describe 
particles and antiparticles without electric charge (cf. Landau and Lifchitz \cite{Landau}). In this case the conserved 
charge $Q$ is associated with the difference between the numbers of particles and antiparticles.

The lagrangian density is:  
\be
\label{2.16}
{\cal L}=-\partial_{\mu}\phi^{*}\partial^{\mu}\phi-V(\phi^{*}\phi)
\ee
with potential

\be
\label{2.17}
V(\phi^{*}\phi)=m^2 \phi^{*}\phi+\lambda(\phi^{*}\phi)^2,
\ee
The constant $m^2$ is positive and represents the squared mass of the 
particles, the coupling constant $\lambda$ is also positive so that the 
interaction is repulsive. If $\lambda$ is small we can describe an almost 
ideal relativistic gas of bosons at temperature $T=0$ by the field 
\be
\label{2.14}
\hat{\psi}=\phi+\hat{\psi}'
\ee
where $\phi$ is a classical field which will represent the condensate; it 
verifies the Euler-Lagrange equation
\be
\label{2.18}
-\partial_{\mu}\partial^{\mu}\phi+m^2\phi+2\lambda\phi^{*}\phi\phi=0
\ee
and $\hat{\psi}'$ is a fluctuation.

If we suppose that the system is enclosed in a box of volume $V$, the field
can be expanded in a Fourier series
\be
\label{2.1}
\hat{\psi} (x)=\sum_{\bf p}\frac{1}{\sqrt{V}\sqrt{2\epsilon}}(a_{\bf p}
e^{ipx}+b^{+}_{\bf p}e^{-ipx}),
\ee
with$$px=-p^{0}t+{\bf p}\cdot{\bf x}$$ and

\be
\label{2.2}
\epsilon=p^{0}=\sqrt{{\bf p}^{2}+\epsilon_0^{2}},
\ee
where $\epsilon_{0}$ is a constant whose precise value will be
fixed below. The operators $a_{\bf p},b_{\bf p}$ and their conjugates 
$a^{+}_{\bf p},b^{+}_{\bf p}$ are the annihilation and creation operators for 
the particles $a$ and antiparticles $b$:

\be
\label{2.3}
[a_{\bf p},a^{+}_{\bf p}]=1, \; [b_{\bf p},b^{+}_{\bf p}]=1.
\ee

Let us note that the Fourier expansion (\ref{2.1}) and 
the commutation relations (\ref{2.3}) usually given for free fields are valid also for 
interacting fields (cf. Lee \cite{lee}) .
Of course in this case $a_{\bf p}$ and $b_{\bf p}$ and conjugates can have a 
complicated time dependence in contrast to free fields where these 
operators are time independant if we choose for $\epsilon_{0}$ the mass
$m$ of the free particle.

For a uniform weakly non ideal gas in a large volume $V$ 
at temperature $T=0$, almost all the $N$ particles, $N_0\approx N$, 
condense in the zero mode of the particles. As it will be justified 
in the following section, this mode represents a minimum of energy 
when we impose the constraint that the conserved charge $Q$ is 
fixed: $Q=N_0$. Hence, since $N_0$ is a macroscopic number, following 
the Bogoliubov approximation \cite{bog47}, we can negect the commutator $[a_{\bf 0},a^{+}_{\bf 0}]=1$ as compared with 
the huge eigenvalue $N_0$ of $a_{\bf 0}^{+}a_{\bf 0}$, so the operators $a_{\bf 0}$, $a_{\bf 0}^{+}$  can 
be approximated by classical numbers  
\be
\label{4'}
a_{\bf 0}\approx a_{\bf 0}^{+}\approx\sqrt{N_0}.
\ee
Therefore the Fourier expansion (\ref{2.1}) can be decomposed
\be
\label{2.11}
\hat{\psi}=\frac{\sqrt{N_0}}{\sqrt{V}\sqrt{2\epsilon_0}}e^{-i\epsilon_{0}t}+
\hat{\psi}'
\ee
where the classical field
\be
\label{2.12}
\phi_0=\frac{\sqrt{N_0}}{\sqrt{V}\sqrt{2\epsilon_0}}e^{-i\epsilon_{0}t} 
\ee
represents the condensate of the particles of zero momentum and the operator 
$\hat{\psi}'$ referring to the remaining non condensed particles is 
considered as a small perturbation.

Let us note that we have not fixed the arbitrary constant $\epsilon_0$. 
Since $\phi_0$ given by (\ref{2.12}) must be a classical solution of 
the equation (\ref{2.18}), $\epsilon_0$, supposed positive, is fixed by
\be
\label{2.13}
n_0=\frac{\epsilon_{0}(\epsilon_0^2-m^2)}{\lambda}
\ee
where
\be
\label{2.12'}
n_0=\frac{N_0}{V}
\ee
is the constant density.

Let us remark that we consider a gas of bosons at low energy. 
More precisely it is a relativistic gas but a weakly relativistic gas where creation and 
destruction of antibosons are negligible. Of course we could also 
consider the symmetrical situation of a weakly relativistic gas of 
antibosons living in an antimatter world. In this case the constraint is $Q=-N_0$ 
where $N_0$ is the number of antibosons, the 
condensation would be on the zero mode of the antiparticles and the classical 
field representing the antiparticle condensate would be given by 
(\ref{2.12}) with opposite phase.

The generalisation of the Bogoliubov approximation to the case of non 
uniform and time dependent configurations is given by (\ref{2.14})
where the classical field $\phi$, which represents the condensate, verifies the equation (\ref{2.18}).
This completes the comparison with the non relativistic theory (cf. for example \cite{dalf}).

%%%%%%%%%%%%%%%%%%%%%%%%%%%%%%%%%%%%%%%%%%%%%%%%%%%%%%%%%%%

\section{Minimum of energy at fixed conserved charge}

%%%%%%%%%%%%%%%%%%%%%%%%%%%%%%%%%%%%%%%%%%%%%%%%%%%%%%%%%%%

In this section we show that the homogeneous solution (\ref{2.12}) and 
(\ref{2.13})  realises the minimum of energy when 
the number of particles is fixed.

It will be convenient to write the classical complex field $\phi (x)$ 
in polar form:

\be
\label{3'.1}
\phi (x)=F(x)e^{iS(x)}.
\ee
The Lagrangian (\ref{2.16}) becomes
\be
\label{lag}
{\cal L}=-\partial_{\mu}F\partial^{\mu}F
-F^{2}\partial_{\mu}S\partial^{\mu}S-V(F^{2})
\ee
and the Euler-Lagrange equations take the form
\be
\label{3'.2}
-\frac{\nabla_{\mu}\nabla^{\mu}F}{F}+\partial_{\mu}S
\partial^{\mu}S+V'(F^{2})=0
\ee
where $V'(F^{2})$ is the derivative $\frac{dV}{dF^2}$,
\be
\label{3'.3}
\nabla_{\mu} (F^{2}\partial^{\mu}S)=0.
\ee
The 4-current and the energy density are given respectively by
\be
\label{3'.4}
j_\mu=-i(\phi^{*}\partial_{\mu}\phi-\partial_{\mu}\phi^{*}\phi)
=2F^{2}\partial_{\mu}S
\ee
and
\be
\label{3'.6}
{\cal H}=(\partial_{0} F)^{2}+F^{2}(\partial_{0} S)^{2}+(\nabla F)^{2}
+F^{2}(\nabla S)^{2}+V(F^{2}).
\ee
Let us introduce the conjugate momenta of the fields
\be
\label{momf}
\pi_{F}=\frac{\partial {\cal L}}{\partial F_{,0}}=2\partial _{0}F
\ee
and
\be
\label{moms}
\pi_{S}=\frac{\partial {\cal L}}{\partial S_{,0}}=2F^{2}\partial _{0}S.
\ee

The charge and the Hamiltonian can be respectively written 
\be
\label{3'.5}
Q=\int_{V}j^{0}d^{3}x=\int_{V}-2F^{2}\partial_{0}Sd^{3}x =
\int_{V}-\pi_{S}d^{3}x
\ee
and
\be
\label{3'.7}
H= \int_{V}(\frac{1}{4}\pi_{F}^{2} 
+\frac{1}{4}\frac{\pi_{S}^{2}}{F^{2}}+(\nabla F)^{2} 
+F^{2}(\nabla S)^{2}+V(F^{2}))d^{3}x.
\ee

The extremum of the energy $H$, when the conserved charge $Q$ is fixed by 
the constant $N_0$, is determined by the variation
\be
\label{variation}
\delta\left(H-\mu (Q-N_{0})\right)=0
\ee
where $\mu$ is a Lagrange multiplier.
The variation on  $\pi_{F}$, $\pi_{S}$, $F$, $S$, gives respectively
\be
\label{3.12}
\pi_{F}=0,
\ee
\be
\label{3'.13}
\frac{1}{2}\frac{\pi_{S}}{F^{2}}+\mu =0,
\ee
that is
\be
\label{3'.14}
F=F({\bf x}),
\ee
\be
\label{3'.15}
S= -\mu t+S_{0}({\bf x}),
\ee
and implies that these functions are solutions of the equations (\ref{3'.2}), (\ref{3'.3}).

Bekenstein and Guendelman \cite{bek-guen} show that this type of solution is a local energy minimum 
within its sector of fixed charge. 
In particular the homogeneous solution (\ref{2.12}) and 
(\ref{2.13})  belongs to this class of solutions with $\mu=\epsilon_{0}$
and realises the minimum of energy when 
the conserved charge is fixed by the number $N_0$ of particles, that is
\be
\label{min}
H=\frac{3}{4} N_{0}\epsilon_{0}+\frac{1}{4} N_{0}\frac{m^{2}}{\epsilon_{0}}.
\ee

We can show that (\ref{min}) is effectively the energy minimum as follows.
A solution of (\ref{variation}) can be written as a perturbation of the
homogeneous solution (\ref{2.12}), (\ref{2.13}):
\be
\label{dev1}
F=\sqrt{\frac{n_0}{2\epsilon_0}}+f({\bf x}), 
\ee

\be
\label{dev2}
S=-\epsilon_{0}t+S_{0}({\bf x}) 
\ee

where $f({\bf x})$ is constrained by
\be
\label{contrainte}
\int_{V}\left(2\sqrt{\frac{n_0}{2\epsilon_0}}f({\bf x})
+f^{2}({\bf x})\right)d^{3}x=0
\ee
in order to maintain $Q=N_0$.

If we introduce (\ref{dev1}) and (\ref{dev2}) in  (\ref{3'.7}), 
several terms are canceled by using (\ref{contrainte}) and we obtain finally
\ba
\label{devmin} 
\nonumber & & H=\frac{3}{4} N_{0}\epsilon_{0}+\frac{1}{4} N_{0}\frac{m^{2}}{\epsilon_{0}}\\ 
& & +\int_{V}\left((\nabla f({\bf x}))^{2}
+(\sqrt{\frac{n_0}{2\epsilon_0}}+f({\bf x}))^{2}(\nabla S_{0}({\bf x}))^{2}
+\lambda \left(2\sqrt{\frac{n_0}{2\epsilon_0}}f({\bf x})
+f^{2}({\bf x})\right)^{2}\right)d^{3}x
\ea
The integral on the right hand side is obviously positive therefore (\ref{min}) 
is the minimum with $Q=N_0$.

%%%%%%%%%%%%%%%%%%%%%%%%%%%%%%%%%%%%%%%%%%%%%%%%%%%%%%%%%%%%%%%%%

\section{Fluid description}

%%%%%%%%%%%%%%%%%%%%%%%%%%%%%%%%%%%%%%%%%%%%%%%%%%%%%%%%%%%%%%%%%

We shall express the modulus $F(x)$ of the field in the form
\be
\label{3.2}
F^{2}(x)=\frac{n(x)}{2h(x)}.
\ee
Below it will turn out that $n(x)$ is the proper density of the particles
and $h(x)$ the enthalpy per particle in the fluid approximation. 
The expression (\ref{3.2}) can be suggested by the homogeneous solution 
(\ref{2.12}) where $\frac{N_0}{V}=n_0$ is the constant density and 
$\epsilon _{0}$ an energy scale fixed by the relation 
(\ref{2.13}).

The 4-gradient of the phase $\partial_{\mu}S$ can be written as a 
covector $C_\mu$ in terms of a unit 4-vector $u^\mu$:
\be
\label{3.6}
C_\mu=\partial_{\mu}S=hu_\mu
\ee
hence from (\ref{3.2}) the 4-current (\ref{3'.4}) takes the hydrodynamics form:

\be
\label{3.7}
j_\mu=nu_\mu
\ee
where we interpret $n$ (if it is positive) as the proper density and $u^\mu$ 
as the velocity 4-vector of the fluid.

From (\ref{3.6}) we deduce also

\be
\label{3.8}
\partial_{\mu}C_\nu-\partial_{\nu}C_\mu=0
\ee
and
\be
\label{3.9}
C_{\mu}C^\mu=-h^2
\ee

Equation (\ref{3'.3}) is the familiar continuity equation
\be
\label{3.13}
\nabla_{\mu}j^\mu=0.
\ee
In equation (\ref{3'.2}) the term $\frac{\nabla_{\mu}\nabla^{\mu}F}{F}$ 
can be neglected if $F$ varies slowly compared to $S$. This term is analogous to 
the ``quantum pressure'' $\frac{\hbar^2}{2m}\frac{\nabla^{2}\sqrt{n}}
{\sqrt{n}}$ in the non relativistic theory \cite{dalf}.

Let us look at this approximation more closely. The equation
(\ref{3'.2}) takes the form

\be
\label{3.14}
\partial_{\mu}S\partial^{\mu}S+V'(F^2)=0,
\ee
hence we have

\be
\label{3.15}
-C^{\mu}C_\mu=m^2+\lambda\frac{n}{h}
\ee
which combined with (\ref{3.9}) gives
\be
\label{3.16}
n=\frac{h(h^2-m^2)}{\lambda}.
\ee
This relation $n(h)$ is identical with the relation $n_0(\epsilon_0)$
given by (\ref{2.13}), but (\ref{3.16}) is valid within the approximation 
considered here.

The equations (\ref{3.8}),(\ref{3.9}),(\ref{3.13}) and
\be
\label{3.17}
j_\mu=\frac{n(h)}{h}C_\mu
\ee
are the relativistic equations of a perfect isentropic irrotational fluid 
( cf. \cite{lich} and \cite{bb} where these equations appear in the same notations). For the most general framework 
in relativistic hydrodynamics (multicomponent fluids, superfluids), the reader is refered to \cite{carterfluid} 
and \cite{carlan98}.)

The nature of $h$ is now clear. It is the enthalpy per 
particle (with the mass energy $m$ included). The equation 
(\ref{3.16}) is the equation of state which is automatically imposed.

If the quantities $\rho$ and $p$ designate respectively the proper 
energy density and the pressure, the enthalpy per particle can be written
\be
\label{3.18}
h=\frac{\rho+p}{n}.
\ee
Since the entropy is zero we have
\be
\label{3.19}
dp=ndh
\ee
The equations (\ref{3.18}) and (\ref{3.19}) define $p$ and $\rho$ in function 
of $h$ and $n$. (\ref{3.19}) combined with (\ref{3.16}) can be integrated:

\be
\label{3.20}
p=\frac{(h^2-m^2)^2}{4\lambda}=\lambda F^4 (x).
\ee
The velocity of sound $v_{s}=\sqrt{dp/d\rho}$ is given by 

\be
\label{3.21}
v_{s}^{2}=\frac{n(h)}{hn'(h)}=\frac{h^2-m^2}{3h^2-m^2} < \frac{1}{3}.
\ee
We remark that $v_{s}^{2} \rightarrow 1/3$ when $m \rightarrow 0$ which is the 
the correct ultrarelativistic limit.

The expressions(\ref{3.18}) and (\ref{3.20}) can also be obtained by 
identifying the energy-momentum tensor of (\ref{lag}) (in which 
the derivatives $\partial_{\mu} F$ are neglected), that is

\be
\label{3.22}
T_{\mu\nu}=2F^{2}\partial_{\mu}S\partial_{\nu}S+\eta_{\mu\nu}
(-F^{2}\partial_{\rho}S\partial^{\rho}S-V(F^{2}))
\ee
with the perfect fluid energy-momentum tensor.
Finally we notice that the approximate equations can be obtained from
a generic Lagrangian proposed by Carter \cite{houches} 
to describe the irrotational fluid in a condensate:

\be
\label{cart}
{\cal L}=-\frac{1}{2}F^{2}\partial_{\mu}S\partial^{\mu}S- {\cal U}(F).
\ee
In our case:

\be
\label{3.23}
\begin{array}{lcr}
{\cal U}(F) =\frac{1}{2}(m^{2}F^2+\lambda F^4). 
\end{array}
\ee

%%%%%%%%%%%%%%%%%%%%%%%%%%%%%%%%%%%%%%%%%%%%%%%%%%%%%%%%%%

\section{Cylindrical vortex}

%%%%%%%%%%%%%%%%%%%%%%%%%%%%%%%%%%%%%%%%%%%%%%%%%%%%%%%%%%

\subsection{Vortex solution of the fluid approximation}

It is convenient to begin with the approximate theory governed 
by the equations 
(\ref{3.8}), (\ref{3.9}), (\ref{3.13}) and (\ref{3.17}) with the equation of
state of the fluid (\ref{3.16}).

We know (cf. \cite{bb}, \cite{carlan52}, \cite{prix}) that in cylindrical coordinates

\be
\label{4.1}
ds^2=-dt^2+dz^2+dr^2+r^2d\theta^2
\ee
a stationary cylindrical solution is given by 

\be
\label{4.2}
C_\mu=(-E , L , 0 , M)
\ee
where $E$, $L$, $M$ are the three constants associated with the three 
symmetries. By a Lorentz transformation along $z$ we can always choose $L=0$ 
and have a circular flow.

The enthalpy $h$ is a function of $r$ given by

\be
\label{4.3}
h^2=-C^\mu C_\mu=E^2-L^2-\frac{M^2}{r^2}.
\ee

The equation of state of the fluid (\ref{3.16}) yields two values $h=0$ and 
$h=m$ for which $n$ vanishes. When we move from the exterior toward the center 
of the vortex, $h(r)$ is decreasing; $h$ is always larger than $m$, so 
$n$ vanishes first at $h=m$ which determines the radius 

$$r_0=\frac{M}{\sqrt{E^2-L^2-m^2}}$$
of the core of the vortex. Below this value $n$ becomes negative, which 
is meaningless in this approximation of classical fluid. Let us 
observe that with a familiar polytropic fluid the core is 
empty, $n$ is zero. The sound velocity vanishes also on the radius $r_0$. 

We shall now suppose for simplicity that we have a circular flow with 
$L=0$. Using (\ref{3.16}), (\ref{4.2}), (\ref{4.3}), we write the wave 
function of the vortex of the approximate equations:

\be
\label{4.4}
\phi= \left(\frac{E^2-\frac{M^2}{r^2}-m^2}{2\lambda}\right)^{1/2} 
e^{i(-Et+M\theta)}
\ee
where we have suppressed an undetermined constant phase. 

To make this wave function single-valued, $M$ must be an integer, noted 
$\nu$. The domain of validity of (\ref{4.4}) is $r\geq r_0$ where 

\be
\label{4.5}
r_0=\frac{1}{\sqrt{E^2-m^2}}
\ee
is the core radius for $\nu=1$.
The length $r_0$ sets the limit below which we cannot neglect 
the effect of the quantum term. 

It is immediate to write the velocity of the 
circular flow  $$v=\frac{\nu}{Er}.$$
Let $v_0$ be this velocity  at $r=r_0$ for $\nu=1$, we have:
\be
\label{4.51}
E=\frac{m}{\sqrt{1-v_0^2}}.
\ee
%%%%%%%%%%%%%%%%%%%%%%%%%%%%%%%%%%%%%%%%%%%%%%
\subsection{Vortex solution of the exact theory}
%%%%%%%%%%%%%%%%%%%%%%%%%%%%%%%%%%%%%%%%%%%%%%

A vortex solution of the exact equations (\ref{3'.2}) and (\ref{3'.3}) 
must verify (\ref{3.8}), (\ref{3.9}), (\ref{3.13}) and (\ref{3.17}) 
whose stationary cylindrical solution is always given by (\ref{4.2}) 
and (\ref{4.3}). The relation between $n$ and $h$ is no longer (\ref{3.16}) 
but $n$ is a function of $r$ and $F^{2}=\frac{n(r)}{2h(r)}$ is also a 
function of $r$. So the vortex $\phi=F\exp{iS}$ is 
determined (with $L=0$) by 
\be
\label{4.6}
S=-Et+\nu \theta
\ee
and by the the differential equation for $F(r)$ deduced from 
(\ref{3'.2}):

\be
\label{4.7}
\frac{1}{r}\partial_{r}(r\partial_{r}F(r))-\frac{\nu^2}{r^2}F(r)+(E^2-m^2)F(r)
-2\lambda F^{3}(r)=0
\ee   

This type of equation has been discussed in the literature many times (see 
for example \cite{gros63}). Analytical solutions are not kown, but it is easy 
to discuss its asymptotic properties for $r\rightarrow 0$ and 
$r\rightarrow \infty$ and to compare with the solution of the 
fluid approximation.
The presence of the term $\nu^2F/r^2$ means that $F(r)$ must tend to 
zero when $r\rightarrow 0$ if the energy in the volume $V$ has to be finite. 
For small $r$ the cubic term $2\lambda F^3$ is the smallest and can be 
neglected, the solution is the Bessel function

\be
\label{4.8}
F(r)=A J_\nu[(E^2-m^2)^{1/2} r]
\ee
as $r\rightarrow 0$.

As $r\rightarrow \infty$ the first two terms of (\ref{4.7}) cancel:

\be
\label{4.9}
F(r)= F_0=\sqrt{\frac{E^2-m^2}{2\lambda}}.
\ee

An approximation to the solution is obtained by matching the functions 
(\ref{4.8}) and (\ref{4.9}) and their derivatives. Let $r=a_0$ be the radius 
of the junction. The constant A is fixed by the continuity of $F$ at
$r=a_0$ (we are not concerned by this value) and $a_0$ by the continuity of 
the derivative $\frac{dF}{dr}(a_0)$:

\be
\label{4.10}
J'_{\nu}[(E^2-m^2)^{1/2} a_0]=0.
\ee
So $a_0$ is given in term of the first zero of $J'_{\nu}$, 
let say $z_\nu$. 
For the winding number $\nu=1$, $z_1=1.84$.
The radius $a_0$ of the junction, which is naturally interpreted as the 
radius of the vortex, is given by
\be
\label{4.11}
a_0=\frac{1.84}{\sqrt{E^2-m^2}}.
\ee

We observe that $r_0$ the radius of the core of the vortex in the fluid 
approximation and $a_0$ differ only in a numerical factor close to unity, 
they are typically of the same order of magnitude. The length scale 
$a_0 \sim r_0$ is of the order of the so called ``healing length'' 
\cite{dalf}. This result indicates that the fluid approximation 
is a good practical model for the study of the vortices 
in a relativistic Bose gas condensate. 

%%%%%%%%%%%%%%%%%%%%%%%%%%%%%%%%%%%%%%%%%%%%%%

\section{Kelvin waves on the cylindrical vortex}

%%%%%%%%%%%%%%%%%%%%%%%%%%%%%%%%%%%%%%%%%%%%%%

We consider now a generic perturbation of the vortex $\phi=F\exp{(iS)}$ determined by (\ref{4.6}) and (\ref{4.7}):

\be
\label{105}
\tilde{\phi}=\phi+\Delta \phi \simeq e^{iS}(F+\Delta F+iF\Delta S).
\ee
Substituting  (\ref{105}) in the Euler-Lagrange equations and assuming 
that the perturbation is small, we can linearise in $\Delta F$ andin $F\Delta S$. 
Then it is possible to expand $\Delta F$ and $F\Delta S$ in wave modes and to consider each mode separately:
 \be
\label{106}
\Delta F=M(r)\cos (-\omega t+qz+\bar{\nu}\theta),
\ee
 \be
\label{107}
F\Delta S=N(r)\sin (-\omega t+qz+\bar{\nu}\theta).
\ee
$M(r)$ and $N(r)$ verify the linear equations
\be
\label{108}
M''+\frac{1}{r} M'+(E^2+\omega^2-q^2-6\lambda F^2-m^2)M-\frac{1}{r^2}[(\bar{\nu}^2+\nu^2)M+2\bar{\nu}\nu N]+2E\omega N=0,
\ee
\be
\label{109}
N''+\frac{1}{r} N'+(E^2+\omega^2-q^2-2\lambda F^2-m^2)N-\frac{1}{r^2}[(\bar{\nu}^2+\nu^2)N+2\bar{\nu}\nu M]+2E\omega M=0,
\ee

We can also \cite{ben4} introduce equivalently
\be
\label{110}
\Delta F + iF\Delta S= H_{+}(r)\exp{ i (-\omega t+qz+\bar{\nu}\theta)}+ H_{-}(r)\exp{ -i(-\omega t+qz+\bar{\nu}\theta)}
\ee
with
\be
\label{111}
H_{+}(r)=\frac{M(r)+N(r)}{2} \quad ,\quad  H_{-}(r)=\frac{M(r)-N(r)}{2}
\ee
which verify the linear equations:
\be
\label{112}
\frac{1}{r}(rH_{\pm}')' + [(E \pm \omega)^2 -q^2- \frac{(\nu\pm\bar{\nu})^2}{r^2}]H_{\pm}= (m^2+4\lambda F^2)H_{\pm}+2\lambda F^2 H_{\mp}
\ee
Since we are interested in the small motion of the central line of the vortex, let us concetrate in the neighbourhood of the origine.
For small $r$, $F(r)$ is given by (\ref{4.8}) hence
\be
\label{114}
F(r) \sim A r^\nu.
\ee
We can neglect  the two last termes of the r.h.s. of Eqs.(\ref{112}) which are approximed by Bessel equations whose the solutions are 
\be
\label{115}
H_{\pm}= B J_{\vert \nu \pm \bar{\nu}\vert}[((E\pm \omega)^2-q^2-m^2)^{\frac{1}{2}}r] \propto r^{\vert \nu \pm \bar{\nu}\vert}  
\ee
The center of the perturbed vortex is determined by the zero of the solution $\tilde{\phi}$ and since $F(r)$ vanishes at 
the origin, Eq.(\ref{105}) requires that at least one of the functions $H_{\pm}$ is different of zero at the origin. 
Suppose it is $H_{+}$, we must have
\be
\label{116}
\nu + \bar{\nu}=0
\ee
and consequently $H_{-} \propto r^{2\nu}$ is negligible. Hence near the origin we get
\be
\label{117}
\tilde{\phi} = \exp{i(-Et+\nu \theta)}\left(Ar^\nu  +H_{+}(r) \exp{i(-\omega t+qz+\bar{\nu}\theta)}\right). 
\ee
Let the constant $B =-a^{\nu}A$, eq. (\ref{117}) becomes 
\be
\label{118}
\tilde{\phi} = A\exp{i(-Et)}\left(r^\nu e^{i\nu\theta}  -a^{\nu} \exp{i(-\omega t+qz)}\right). 
\ee
In the case $\nu=1$ the cancellation of $\tilde{\phi}$ is obtained for
\be
\label{119}
r=a \quad ,\quad \theta= -\omega t +qz
\ee
which is the helical motion of the center of the vortex, that is the Kelvin waves.

Let us return to the equations (\ref{108}) and (\ref{109}). If $E=0$ and if we change $m^2$ into $-m^2$, 
they are identical to the corresponding equations 
for the global strings of \cite{gradw}. We can verify that for $\nu=1$, $\bar{\nu}=-1$, $\omega=0$ and $q=0$ there is 
an exact solution:
\be
\label{120}
M(r)=F'(r) \quad , \quad N(r)=\frac{F(r)}{r}.
\ee
In the following we shall  have always $\nu=-\bar{\nu}=1$ since we are interested 
in Kelvin waves and we shall use non dimensional 
variables.

Let 
\be
\label{121}
y=\frac{r}{r_0},
\ee
\be
\label{122}
F(r) = F_0 f(\frac{r}{r_0})=F_0 f(y)
\ee  
where $r_0$ and $F_0$ are given respectively by (\ref{4.5}) and (\ref{4.9}).

The equation (\ref{4.7}) becomes the standard equation 
\be
\label{123}
f'' +\frac{1}{y}f'+(1-\frac{1}{y^2})f-f^3 =0.
\ee
For large $y$ the solution is
\be
\label{124}
f(y)=1-\frac{1}{2y^2}-\frac{9}{8y^4}-\frac{161}{16y^6}+\cdots,
\ee
for small $y$ we have
\be
\label{125}
f(y)= ky+O(y^3)
\ee
where $k$ is a constant which can be obtained numerically.

From the equations (\ref{106}) and (\ref{107}) we have naturally
\be
\label{126}
M(r)=F_0\bar{M}(y) \quad and \quad N(r)= F_0 \bar{N}(y).
\ee
We introduce also $\bar{\omega}=\omega r_0$, $\bar{q}=q r_0$, $\bar{E}=E r_0$, $\bar{m}=m r_0$, $t=r_0\bar{t}$, $z=r_0\bar{z}$.
The equations (\ref{108}) and (\ref{109}) becomes:
\be
\label{127}
\bar{M}''+\frac{1}{y}\bar{M}'+(\bar{\omega}^2- \bar{q}^2+1-3f^2(y))\bar{M}-\frac{2}{y^2}(\bar{M}-\bar{N})+2\bar{E}\bar{\omega}\bar{N}=0,
\ee
\be
\label{128}
\bar{N}''+\frac{1}{y}\bar{N}'+(\bar{\omega}^2- \bar{q}^2+1-f^2(y))\bar{N}-\frac{2}{y^2}(\bar{N}-\bar{M})+2\bar{E}\bar{\omega}\bar{M}=0
\ee
Since $\nu=-\bar{\nu}=1$ the solution of these equations can express the deplacement of the vortex. This is the case 
of the exact solution:
\be
\label{129}
\bar{M}_0=f'(y) \quad , \quad \bar{N}_0=\frac{f(y)}{y}
\ee
obtained for $\bar{\omega}=\bar{q}=0$.

We can now examine the solutions of Eqs. (\ref{127}) and (\ref{128}) and compute 
the frequency $\bar{\omega}$ in the limit
$\bar{q} \rightarrow 0$ of the long wavelength oscillations.
To solve the equations we shall follow the method of ``matched asymptotic expansions'' used by 
Roberts and Grant \cite{robertsgrant,grant}.

We consider first the solution in the inner cylindrical region of radius $y$ centred on the $oz$ axis in 
the form of an expansion about
the exact solution (\ref{129}). This solution is then examined for $y \rightarrow \infty$. 
We consider also the solution in the outer 
region caracterized 
by a new stretched coordinate $s=\bar{q} y$ which is of order unity for large $y$.
This solution is examined in the limit $s \rightarrow 0$. Finally the two asymptotic 
solutions ($y \rightarrow \infty$ and 
$s \rightarrow 0$) are matched in an overlap 
domain, where both the inner and outer expansion are valid. 

From this matching we shall obtain the dispersion relation of the Kelvin waves.

%%%%%%%%%%%%%%%%%%%%%%%%%%%%%%%%%%%%%%%%%%%%%%%
\subsection{Interior solution}
%%%%%%%%%%%%%%%%%%%%%%%%%%%%%%%%%%%%%%%%%%%%%%
We expand $\bar{M}$ and $\bar{N}$ around the exact solution $\bar{M}_0$ and $\bar{N}_0$:
\be
\label{130}
\bar{M}=\bar{M}_0+\bar{q}^2 \bar{M}_1+\cdots
\ee
\be
\label{131}
\bar{N}=\bar{N}_0+\bar{q}^2 \bar{N}_1+\cdots.
\ee
Then, substituting (\ref{130}) and (\ref{131}) into the equations (\ref{127}) and (\ref{128}) we obtain, by equating
the coefficients of $\bar{q}^2$
\be
\label{132}
\bar{M}_1''+\frac{1}{y}\bar{M}_1'+\left(1-\frac{2}{y^2}-3f^2(y)\right)\bar{M}_1+\frac{2}{y^2}\bar{N}_1= 
f'(y)-2\bar{E}\omega_1\frac{f(y)}{y},
\ee
\be
\label{133}
\bar{N}_1''+\frac{1}{y}\bar{N}_1'+\left(1-\frac{2}{y^2}-f^2(y)\right)\bar{N}_1+\frac{2}{y^2}\bar{M}_1= 
\frac{f(y)}{y}-2\bar{E}\omega_1f'(y),
\ee
with
\be
\label{134}
\bar{\omega}=\omega_1 \bar{q}^2.
\ee
To solve (\ref{132}) and (\ref{133}) we substitute $f(y)$ by the expansion (\ref{124}) valid for large $y$.

First we search for solutions for the homogeneous system which is transformed in a linear equation of $4^{th}$ order in $\bar{M}_1$:
\ba
\label{135}
\nonumber & & \bar{M}_1^{(4)}+\frac{6}{y}\bar{M}_1^{(3)}+(-2+\frac{7}{y^2}+\frac{8}{y^4}+\frac{76}{y^6} +\cdots)\bar{M}_1''
+(-\frac{10}{y}+\frac{1}{y^3}-\frac{16}{y^5}-\frac{380}{y^7} +\cdots)\bar{M}_1'\\
& & +(-\frac{6}{y^2}-\frac{9}{y^4}-\frac{18}{y^6} +\cdots)\bar{M}_1=0%%%+\frac{886}{y^8}+\frac{228}{y^{10}}+\frac{1083}{y^{12}}. 
\ea
Infinity is an irregular singular point of rank $0$ of this equation. 
The roots of the characteristic equation are $\pm\sqrt{2}$ and $0$. The root $0$ has a 
multiplicity $2$. By applying standard techniques about asymptotic solutions of linear equations \cite{wasov}, it is possible 
to obtain asymptotic expansion of four independant solutions:
\be
\label{136}
\begin{array}{lcr}
g_1=\frac{1}{y^3}+\frac{9}{2y^5}+\frac{483}{8y^7}+\cdots\\%%%%%%+\frac{74357}{48y^9}\\
g_2=\frac{1}{y}+\frac{2\ln{y}}{y^3}+\frac{101}{36y^3}+\frac{9\ln{y}}{y^5}+\frac{27}{4y^5} +\frac{483\ln{y}}{4y^7} +\cdots \\
%%%+\frac{2467}{32y^7}
g_3=\exp{(-\sqrt{2}y)}\frac{1}{\sqrt{y}}(1-\frac{5}{8\sqrt{2}y}+\frac{65}{256y^2}-\frac{5981}{6144\sqrt{2}y^3}
+\frac{181825}{393216y^4}+\cdots)\\
g_4=\exp{(\sqrt{2}y)}\frac{1}{\sqrt{y}}(1+\frac{5}{8\sqrt{2}y}+\frac{65}{256y^2}+\frac{5981}{6144\sqrt{2}y^3}
+\frac{181825}{393216y^4}+\cdots)\\
\end{array}
\ee
From these solutions $\bar{M}_1$ we deduce the corresponding solutions $\bar{N}_1$:
\be
\label{137}
\begin{array}{lcr}
f_1=\frac{1}{y}-\frac{1}{2y^3}-\frac{9}{8y^5}+\cdots\\
%%% -\frac{109}{48y^7}-\frac{3063487}{48y^9}
f_2=y+ \frac{2\ln{y}}{y}+\frac{65}{36y}-\frac{\ln{y}}{y^3}-\frac{77}{18y^3}-\frac{9\ln{y}}{4y^5} +\cdots\\
%%% -\frac{247}{96y^5}-\frac{12411\ln{y}}{4y^7}-\frac{113501}{96y^7}
f_3= \exp{(-\sqrt{2}y)} \frac{1}{\sqrt{y}} (-\frac{1}{y^2}+\frac{37}{8\sqrt{2}y^3}-\frac{2945}{256y^4}+\cdots)\\
%%%+\frac{2304670093}{25165824\sqrt{2}y^5
f_4= \exp{(\sqrt{2}y)} \frac{1}{\sqrt{y}} (-\frac{1}{y^2}-\frac{37}{8\sqrt{2}y^3}-\frac{2945}{256y^4}+\cdots)\\
%%%-\frac{2304670093}{25165824\sqrt{2}y^5
\end{array}
\ee

From the solutions (\ref{136}) and (\ref{137}) we can construct the fundamental matrix of the homogeneous system. Then by the 
variation-of-constant formula, we obtain the solution of the inhomogeneous system (\ref{132}) and (\ref{133}). This solution 
depends on four arbitrary constants. By choosing two of them we can suppress in the solution the exponentials $\exp{(\pm\sqrt{2}y)}$ 
which are not compatible with the exterior solution. Finally, we obtain:
\be
\label{138} 
\bar{M}_1= \frac{1}{2}\frac{\ln{y}}{y}+\frac{1}{y}(A_1+\bar{E}\omega_1)+\cdots,
\ee
\be
\label{139}
\bar{N}_1= \frac{1}{2}y\ln{y}+A_1y+\frac{1}{2y}\ln^2y+\frac{\ln{y}}{y}\left(2A_1+2\bar{E}\omega_1+\frac{3}{4}\right)+\frac{A_2}{y}+\cdots
\ee
where $A_1$ and $A_2$ are the two remaining constants which can be determined by matching the interior solution and 
the exterior solution.

By combining the equations (\ref{132}) and (\ref{133}) we can obtain an integral relation which will be useful later
in determining $\omega$. 
If we multiply  (\ref{132}) by $yf'(y)$ and (\ref{133}) by $f(y)$ and add the two resulting expressions, using Eq. (\ref{123}) we obtain:
\ba
\label{140}
\nonumber & & \frac{d}{dy}\left(\frac{1}{y}f\bar{N}_1+ f\bar{N}_1'- f'\bar{N}_1+yf'\bar{M}_1'-yf^3\bar{M}_1+yf\bar{M}_1+f'\bar{M}_1 
-\frac{1}{y}f\bar{M}_1\right)=\\  
& & - 2\bar{E}\omega_1(f^2)'+y(f')^2+\frac{1}{y}f^2
\ea
We integrate from $0$ to $\infty$ and we obtain after substituting the value of $\bar{M}_1$, $\bar{N}_1$ 
and $f$ for $y\rightarrow \infty$ and $y\rightarrow 0$:
\be
\label{141}
2A_1 +\frac{1}{2}= -2\bar{E}\omega_1+\left\{ \int_{0}^{\infty} y \left(f'\right)^2dy+ 
\lim_{A\rightarrow \infty}\left(\int_{0}^{A}\frac{1}{y}f^2dy
-\ln A\right)\right\}
\ee
In obtaining this relation, $\ln{\infty}$ has been canceled from the r.h.s. and the l.h.s. If $\bar{E}=1$ the 
relation (\ref{141}) is identical 
to the corresponding non relativistic relation of Grant \cite{grant}. The numerical value of the integral part is known:
\be
\label{142}
\left\{ \int_{0}^{\infty} y \left(f'\right)^2dy+ 
\lim_{A\rightarrow \infty}\left(\int_{0}^{A}\frac{1}{y}f^2dy
-\ln A\right)\right\}=-0.115
\ee
%%%%%%%%%%%%%%%%%%%%%%%%%%%%%%%%%%%%%%%%
\subsection{Exterior solution}
%%%%%%%%%%%%%%%%%%%%%%%%%%%%%%%%%%%%%%%%

Let us return to the equations (\ref{127}) and (\ref{128}) and expand them for large $y$. Replacing $y$ by $\frac{s}{\bar{q}}$ where 
$s$ is a streched coordinate since $\bar{q}$ is small, the equations become:
\ba
\label{143}
\nonumber & &\bar{q}^2\frac{d^2\bar{M}}{ds^2}+ \frac{\bar{q}^2}{s}\frac{d\bar{M}}{ds}
+\left[\omega_1^2\bar{q}^4-\bar{q}^2+1-3\left(1-\frac{\bar{q}^2}{s^2}-\frac{2\bar{q}^4}{s^4}\right)\right]\bar{M}\\
& & -2\frac{\bar{q}^2}{s^2}\left(\bar{M}-\bar{N}\right)+2\bar{E}\omega_1\bar{q}^2\bar{N}=0,
\ea

\ba
\label{144}
\nonumber & &\bar{q}^2\frac{d^2\bar{N}}{ds^2}+ \frac{\bar{q}^2}{s}\frac{d\bar{N}}{ds}
+\left[\omega_1^2\bar{q}^4-\bar{q}^2+1-\left(1-\frac{\bar{q}^2}{s^2}-\frac{2\bar{q}^4}{s^4}\right)\right]\bar{N}\\
& & -2\frac{\bar{q}^2}{s^2}\left(\bar{N}-\bar{M}\right)+2\bar{E}\omega_1\bar{q}^2\bar{M}=0.
\ea
Let us expand $\bar{M}$ and $\bar{N}$ in series of $\bar{q}^2$:
\be
\label{145}
\begin{array}{lcr}
\bar{M}=M_{E1}+\bar{q}^2 M_{E2}+\cdots\\
\bar{N}=N_{E1}+\bar{q}^2 N_{E2}+\bar{q}^4N_{E3}+ \cdots\\
\end{array}
\ee
Substituting into (\ref{143}) and equating successive powers of $\bar{q}^2$ we obtain for the powers zero and two of $\bar{q}$, 
respectively

\be
\label{147}
M_{E1}=0,
\ee

\be
\label{148}
M_{E2}=\left(\frac{1}{s^2}+\omega_1 \bar{E}\right)N_{E1}.
\ee
Substituting (\ref{145}) into (\ref{144}) we find that the equality is identically satisfied at the zero order of $\bar{q}$. 
For the orders $\bar{q}^2$ and $\bar{q}^4$ we obtain, respectively:
\be
\label{149}
N_{E1}''+ \frac{1}{s}N_{E1}' -\left(1+ \frac{1}{s^2}\right)N_{E1}=0,
\ee

\be
\label{150}
N_{E2}''+ \frac{1}{s}N_{E2}' -\left(1+ \frac{1}{s^2}\right)N_{E2}=-\left(\omega_1^2+\frac{2}{s^4}\right)N_{E1}
-\left(\frac{2}{s^2}+2\bar{E}\omega_1\right)M_{E2}.
\ee
Let us note that the coefficient multiplying $N_{E3}$ is zero.

The solution of (\ref{149}) bounded at infinity is the modified Bessel function of the first order $K_1(s)$. Its expansion 
about $s=0$ is:
\be
\label{151}
N_{E1}= C\left\{\frac{1}{s}+\frac{s}{2}\ln{\frac{s}{2}}-\frac{s}{4}(1-2\gamma)+\frac{s^3}{16}\ln{\frac{s}{2}}
-\frac{s^3}{32}\left(\frac{5}{2}-2\gamma\right)+\cdots\right\}
\ee
where $C$ is an arbitrary parameter and $\gamma$ the Euler contant.

From (\ref{148}) we have $M_{E2}$. Now the r.h.s. of Eq. (\ref{150}) is known, since (\ref{151}) is the solution of 
the homogeneous part of the equation  (\ref{150}), 
by the variation-of-constant formula we obtain the solution of (\ref{150}):
\ba
\label{153}
\nonumber & & N_{E2}=C\left\{-\frac{1}{2s^3}+\frac{\ln^2(s/2)}{2s}
+\frac{\ln s}{s}\left(\frac{1}{4}+\gamma+2\bar{E}\omega_1\right)\right.\\
& & \left.+\frac{1}{s}\left(\frac{9}{8}-\frac{\gamma}{4}+\frac{\ln 2}{4}-\frac{\ln^2(2)}{2}\right)+\cdots\right\} 
\ea
From (\ref{145}), (\ref{147}), (\ref{148}), (\ref{151}), and (\ref{153}), we have:
\be
\label{154}
\bar{M}= \bar{q}^2 C\left\{\frac{1}{s^3}+\frac{1}{2s}\ln{\frac{s}{2}}
+\frac{1}{s}\left(-\frac{1}{4}+\frac{\gamma}{2}+\bar{E}\omega_1\right)
+s\ln{\frac{s}{2}}\left(\frac{1}{16}+\frac{\bar{E}\omega_1}{2}\right)\right\},
\ee
\ba
\label{155}
\nonumber & & \bar{N}=C\left\{\frac{1}{s}+\frac{s}{2}\ln\frac{s}{2}-\frac{s}{4}(1-2\gamma)+\frac{s^3}{16}\ln\frac{s}{2}
-\frac{s^3}{32}\left(\frac{5}{2}-2\gamma\right)+\cdots\right\}\\
\nonumber & & +\bar{q}^2 C\left\{-\frac{1}{2s^3}+\frac{1}{2s}\ln^2{\frac{s}{2}}+\frac{\ln s}{s}\left(\frac{1}{4}
+\gamma+2\bar{E}\omega_1\right) \right.\\
& & \left. +\frac{1}{s}\left(\frac{9}{8}-\frac{\gamma}{4}+\frac{\ln 2}{4}-\frac{\ln^2( 2)}{4}\right)+\cdots\right\}.
\ea

%%%%%%%%%%%%%%%%%%%%%%%%%%%%%%%%%%%%%%%%
\subsection{Matching}
%%%%%%%%%%%%%%%%%%%%%%%%%%%%%%%%%%%%%%%%
The interior solution for $y \rightarrow\infty$ must match the exterior one for $s\rightarrow 0$ in an overlap domain. 
For the comparison we express the inner solution in term of $s$:
\be
\label{156}
\bar{M}=\bar{q}^3\left\{\frac{1}{s^3}+\frac{1}{2s}\ln{\frac{s}{2}}
+\frac{1}{s}\left(\frac{\ln 2}{2}-\frac{1}{2}\ln{\bar{q}}+A_1+\bar{E}\omega_1\right)\right\}+\cdots,
\ee
\ba
\label{157}
\nonumber & & \bar{N}=\bar{q}\left\{\frac{1}{s}+\frac{s}{2}\ln\frac{s}{2}+s\left( \frac{\ln 2}{2}-\frac{1}{2}\ln\bar{q}
+A_1\right)\right\}\\ 
\nonumber & & +\bar{q}^3\left\{-\frac{1}{2s^3}+\frac{1}{2s}\ln^2\frac{s}{2}+\frac{\ln s}{s}\left(\ln 2 -\ln\bar{q}
+2A_1+2\bar{E}\omega_1+\frac{3}{4}\right) \right. \\
& & \left.  +\frac{1}{s}\left(-\frac{\ln^22}{2}+\frac{\ln^2\bar{q}}{2}- \ln\bar{q}\left( 2A_1+2\bar{E}\omega_1
+\frac{3}{4}\right)+A_2\right)+\cdots\right\}.
\ea
A comparison of (\ref{154}), (\ref{155}) with (\ref{156}), (\ref{157}) for small $s$ shows that 
\be
\label{158}
C=\bar{q} \quad , \quad 2A_1+\frac{1}{2}=\gamma-\ln\frac{2}{\bar{q}}.
\ee
It is surprising that these two equalities are identical to the corresponding relations of Grant \cite{grant}. In fact, (\ref{158}) 
is satisfied at the order $\bar{q}$ and $\bar{q^3}$ 
for the three first order of $s$ in $\bar{M}$ and $\bar{N}$.

Then using (\ref{134}), (\ref{141}), (\ref{142}) and (\ref{158}) we obtain
\be
\label{159}
\bar{E}\bar{\omega}=\frac{\bar{q}^2}{2}\left(\ln \frac{2}{\bar{q}}-\gamma-0.115\right).
\ee
If we return to the physical variables, we have the dispersion relation:
\be
\label{160'}
\omega= \frac{q^2}{2E}\left(\ln\frac{2}{qr_0}-\gamma-0.115\right),
\ee 
that is, with (\ref{4.51}):
\be
\label{160}
\omega= \sqrt{1-v_0^2}\frac{q^2}{2m}\left(\ln\frac{2}{qr_0}-\gamma-0.115\right),
\ee 
where $r_0$ is the core radius (\ref{4.5}) and $v_0$ the velocity at $r=r_0$ of the relativistic classical fluid.

%%%%%%%%%%%%%%%%%%%%%%%%%%%%%%%%%%%%%%%%%%%%%%%%%%%%%%%%%%%%%%%

\section{Conclusion}

%%%%%%%%%%%%%%%%%%%%%%%%%%%%%%%%%%%%%%%%%%%%%%%%%%%%%%%%%%%%%%%

In the begining of this paper we consider the relations between the quantum nature
of the condensate of an almost ideal relativistic Bose gas and its phenomenological 
properties as a relativistic superfluid. 
The quantum origin of the description
imposes inevitably the equation of state of the fluid (\ref{3.16}) which 
gives an acceptable velocity of sound. Incidently, we note that if we choose a potential with spontaneously 
broken symmetry (by changing $m^2$ into $-m^2$ in (\ref{2.17})) the velocity of sound would always be greater
than $1/\sqrt{3}$.  

The study of a cylindrical vortex has shown that this equation of state
makes sense outside the core of the vortex whose the 
radius is approximately of the order of the healing length which 
indicates the onset of the quantum effects.

Concerning the Kelvin waves, it is surprising that the mathematical calculations are so close 
the non relativistic ones. Yet the field equation differs from the 
Gross-Pitaevskii equation essentially in the second time derivative. 
Consequently the main equations (\ref{127}), (\ref{128})  
are different from the corresponding ones of \cite{grant}. The new dispersion formula (\ref{160}) differs from 
the nonrelativistic one by the relativistic  factor $\sqrt{1-v_0^2}$ where $v_0$ represents the velocity of 
the fluid in $r=r_0$ the radius of the core of the vortex. 

It is interesting to compare with the spinning global strings which behave like vortices in a superfluid medium. 
Ben-Ya'acov \cite{ben4} has given a covariant dispersion relation of the Kelvin waves. In the rest frame 
of the materiel medium, for an unit winding number and $\omega<q$, using (\ref{103'}), it can be express as
\be
\label{161}
\omega= \frac{q^2(1-v_1^2)}{2E}\left(\ln\frac{2}{q\sqrt{1-v_1^2}\delta}-\gamma-0.115\right),
\ee  
where $v_1=\frac{\omega}{q}$ is the velocity along $z$ of the Lorentz frame in which the pertubed vortex 
is an helix  at rest. $E$ is not any more given by (\ref{4.51}) but tends towards the particle mass 
in the non relativistic limit. 

The formula (\ref{160}) and (\ref{161}) are different but, both are compatible at weak energy with 
the Grant formula (\ref{102}). However it is worth stressing that (\ref{160}) is certainly 
the better dispersion relation for a weakly relativistic almost ideal Bose gas. 
The superfluid medium concerned by the dispersion relation (\ref{161}) is unusual as the velocity of sound 
is greater than $1/\sqrt{3}$ 

To conclude we note that relativity does not seem very useful in the laboratory  experiments on the 
condensates but could be of interest in astrophysics.

Acknowledgments

I acknowledge discussions with Hector Giacomini, Bernard Linet and Stam Nicolis.

%%%%%%%gross2%%%%%%%%%%%%%%%%%%%%%%%%gross3%%%%%%%%%%%%%%%%%%%%%%%%%%%%%

\end{document}